\documentclass[a4paper,twocolumn,english,aps,prl,amsmath,showpacs,amssymb]{revtex4-1}
\usepackage[pdftex]{graphicx,hyperref}
\usepackage{verbatim}
\usepackage{amsbsy}
\usepackage{amsmath}
\usepackage{color}
\usepackage{bm}
\usepackage{marvosym}
\usepackage[normalem]{ulem}

\begin{document}

\title{Interaction driven metal-insulator transition in strained graphene}

\author{Ho-Kin Tang,$^{1,2}$ E. Laksono,$^{1,2}$ J. N. B. Rodrigues,$^{1,2}$ P. Sengupta,$^{1,3}$ F. F. Assaad,$^{4}$ and S. Adam$^{1,2,5}$}

\affiliation{$^{1}$ Centre for Advanced 2D Materials, National
  University of Singapore, 6 Science Drive 2, Singapore 117546.}
\affiliation{$^{2}$ Department of Physics, Faculty of Science, National University
  of Singapore, 2 Science Drive 3, Singapore 117542.}
\affiliation{$^{3}$ School of Physical and Mathematical Sciences, Nanyang Technological 
  University, 21 Nanyang Link, Singapore 637371.}
\affiliation{$^{4}$ Institut f\"ur Theoretische Physik und Astrophysik, Universit\"at W\"urzburg, 
  Am Hubland, D-97074 W\"urzburg, Germany.}
\affiliation{$^{5}$ Yale-NUS College, 16 College Ave West, Singapore 138527.}


\date{\today}

\begin{abstract}
  The question of whether electron-electron interactions can drive a
  metal to insulator transition in graphene under realistic
  experimental conditions is addressed. Using three representative
  methods to calculate the effective long-range Coulomb interaction
  between $\pi$-electrons in graphene and solving for the ground state using
  quantum Monte Carlo methods, we argue that without
  strain, graphene remains metallic and changing the substrate from
  SiO$_2$ to suspended samples hardly makes any difference. In
  contrast, applying a rather large -- but experimentally realistic --
  uniform and isotropic strain of about $15\%$ seems to be a promising
  route to making graphene an antiferromagnetic Mott insulator.
\end{abstract}

\maketitle


Over the past decade graphene has established itself as a remarkable new material with superlative 
properties \cite{CastroNeto_RMP:2009,DasSarma_RMP:2011}. However, the early hopes to utilize 
it as a next generation transistor have been dashed mostly because graphene remains metallic 
-- these prototypical Dirac fermions are immune to many of the conventional routes for driving 
two-dimensional electron gases into an insulating state, including, for example, Anderson localization and 
percolation transitions (see e.g. Ref.~\cite{Fuhrer-Adam_Nature:2009}). Other mechanisms 
for opening band-gaps including hydrogenation \cite{Elias_Science:2009}, application of uniaxial 
strain \cite{Ni_ACSNano:2008} and forming 
nanoribbons \cite{Han-Kim_PRL:2007} severely degrade graphene's mobility. Very recently, 
moir\'e heterostuctures using graphene and hexagonal boron nitride have shown evidence of an insulating 
phase \cite{Hunt-Jarillo-Herrero_Science:2013,Ponomarenko-Geim_Nature:2011}, although interpreting these results remains somewhat 
controversial \cite{Jung_NatComm:2015,Amorim-Guinea_arXiv:2015,Guinea-Katsnelson_PRL:2014,DasSarma-Li_PRB:2012}.

In this Letter, we explore a different avenue to make graphene insulating, namely, utilizing the 
electron-electron interactions. Despite much study on the effects of interactions in graphene \cite{Kotov_RMP:2012} 
it is surprising how much still remains to be understood. While it is clear that without any electron-electron 
interactions, graphene should be a semi-metal (SM), and that for very strong interactions it should be 
an insulating anti-ferromagnet (AFM), it remains unclear what one should expect for the real graphene 
material. For example, there are distinct claims in the literature that suspended graphene should be insulating, strongly 
metallic and weakly metallic \cite{Drut_PRL:2009,Schuler_PRL:2013,Ulybyshev_PRL:2014}. 
This discussion could have practical relevance as it could be the basis for a low power Mott-transistor \cite{Nakano_Nature:2012}. 

In this work we explore different ways of controlling the effective strength of electron-electron 
interactions in realistic graphene devices, and propose how one can move around its phase diagram. In particular 
(and in contrast to what is widely assumed to be true \cite{Kotov_RMP:2012,DasSarma_RMP:2011}), we demonstrate 
that it is the non-universal, material-specific and short-range part of the electron-electron interactions 
that plays the dominant role in determining graphene's ground state. More interestingly, we conclude 
that application of isotropic strain is considerably more efficient in approaching the SM-AFM phase 
transition than substrate manipulation, providing a new route for driving the system into the elusive 
Mott insulating phase that has yet to be observed experimentally.

The Hubbard model has served as a versatile paradigm to 
study interacting electrons on a lattice. It is defined as an effective model for electrons 
in partially filled narrow energy bands of a crystal's spectrum.
While the canonical
Hubbard model includes only on-site interactions, the effects of longer range 
interactions are incorporated by a straightforward generalization of the two-body
interaction term, described by the Hamiltonian 
\begin{eqnarray}
  \hat{H} &=& - t \sum_{\langle i j \rangle, \sigma} \big( \hat{c}_{\sigma i}^{\dagger} \hat{c}_{\sigma j} 
  + \textrm{h.c.} \big) + \sum_{i} \hat{n}_{i \uparrow} \mathcal{V}_{i i} \hat{n}_{i \downarrow} \nonumber \\
  && + \frac{1}{2} \sum_{i \neq j} \sum_{\sigma, \sigma'} \hat{n}_{i \sigma} \mathcal{V}_{i j} \hat{n}_{j \sigma'} \,, \label{eq:ExtHubbmodelH}
\end{eqnarray}
where $\hat{c}_{i \sigma}^{\dagger}$ ($\hat{c}_{i \sigma}$) creates (annihilates) an electron of spin $\sigma = \uparrow 
\downarrow$ at position $\mathbf{r}_{i}$ while $\hat{n}_{i \sigma} = \hat{c}_{i \sigma}^{\dagger} \hat{c}_{i \sigma}$ gives 
the density of electrons with spin $\sigma$ at position $\mathbf{r}_{i}$. The nearest neighbor hopping integral 
is identified by $t$, while $\mathcal{V}_{i j}$ stands for the interaction between electrons at sites $i$ and 
$j$. We note that a realistic description of graphene requires the parameters $\mathcal{V}_{i j}$ to be fixed 
in accordance with the spatial profile of the (partially screened) Coulomb interaction $\mathcal{V}(r)$ that 
results from the screening of the bare Coulomb interaction by electrons in energy bands other than the 
$\pi$-bands.
\begin{figure}
  \centering
  \includegraphics[width=0.98\columnwidth]{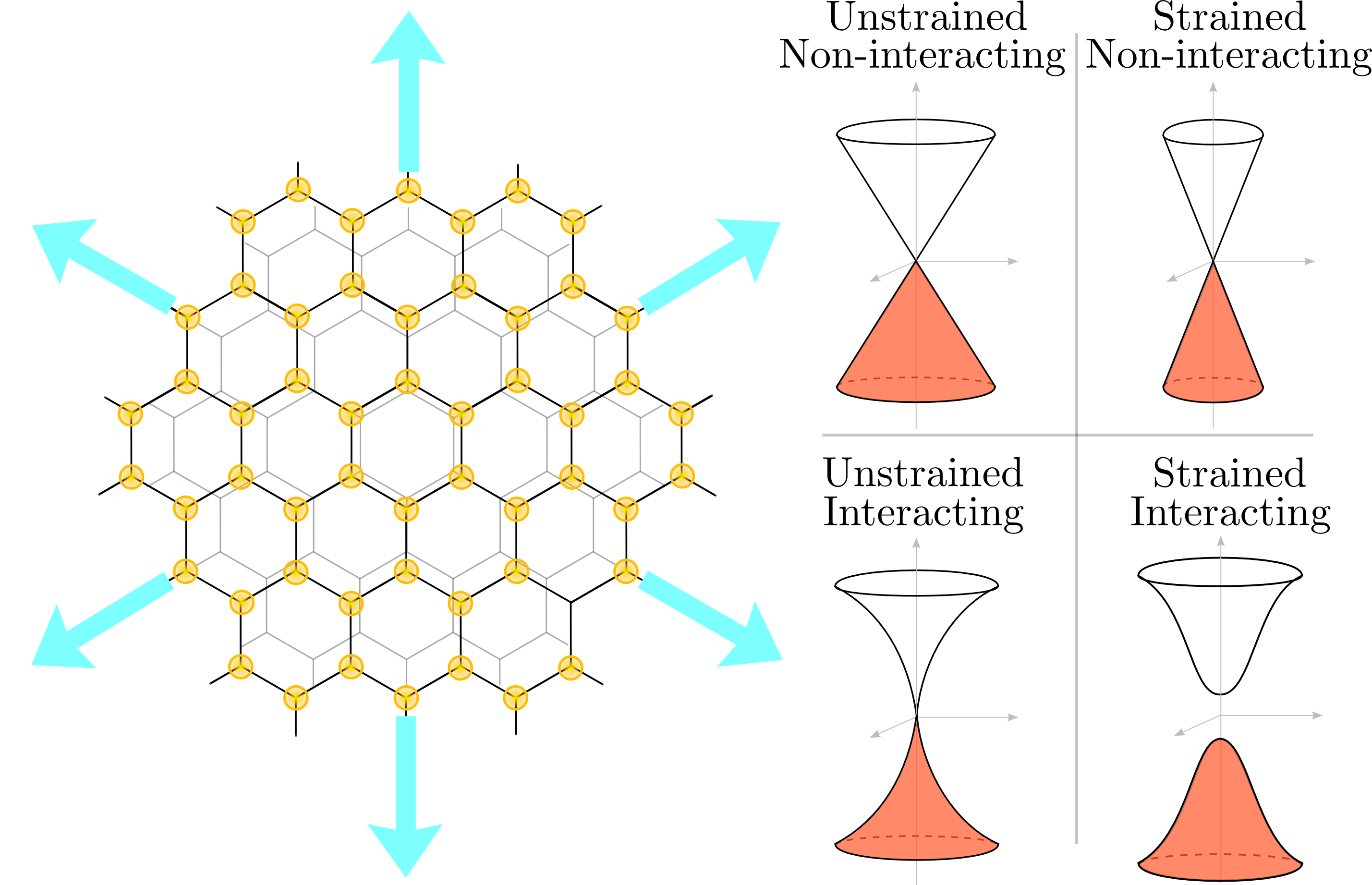}
  \caption{Schematics of biaxially strained graphene (left
    panel). Representation of graphene's low-energy spectrum (right
    panels) of unstrained non-interacting and interacting graphene at
    half-filling, and biaxially strained non-interacting and
    interacting graphene at half-filling.}
  \label{fig:Scheme-Strain}
\end{figure}

It is well established that the canonical Hubbard model [Eq. (\ref{eq:ExtHubbmodelH}) with only on-site interactions, i.e., 
$\mathcal{V}_{i i}=U$ $(>0)$ and $\mathcal{V}_{i j} = 0$ for all $i \neq j$] on the honeycomb lattice at half-filling has a 
critical Hubbard on-site interaction parameter for the SM-AFM transition $U_c = (3.80 \pm 0.01) t$ 
\cite{Sorella_SR:2012,Assaad_PRX:2013,Parisen_PRB:2015}. Following various works based on {\it ab initio} methods (see 
e.g. Ref.~
\cite{Reich_PRB:2002,Jung-MacDonald_WannierPaper_PRB:2013}), it is generally agreed that $t=(2.7 \pm 0.2)~{\rm eV}$. 
Estimates of the on-site interaction parameter $U$ for realistic experimental realizations of graphene vary widely in the literature 
\cite{Yazyev_PRL:2008,Dutta_PRB:2008,Wehling_PRL:2011,Jung_PRB:2011}, with values ranging from $U \approx 1~\rm{eV}$ to 
$10~\rm{eV}$ (where the lower estimates would suggest that graphene is metallic, while the higher estimates hint at it being 
insulating).

However, ignoring longer range interactions in graphene is problematic since the long-range tails of the Coulomb potential 
between Dirac fermions cannot be efficiently screened \cite{Adam_PNAS:2007}.  To address these Coulomb tails, it was recently conjectured 
\cite{Schuler_PRL:2013} that the effects of non-local interactions can be mapped into the Hubbard model with an on-site 
interaction $\widetilde{U}$ given by $\widetilde{U} \approx U - V$, where $U\equiv\mathcal{V}_{i i}$ corresponds to the 
on-site interaction of the long range Hubbard model, while $V\equiv\mathcal{V}_{i\, i+\delta}$ stands for the value of the 
Coulomb potential between electrons at two neighboring atoms on the honeycomb lattice. This effective $\widetilde{U}$ 
would thus be the crucial factor determining graphene's phase. As we discuss below, our numerical calculation with the full long-range potential 
shows that this approximation is qualitatively correct, but quantitatively inaccurate.

\begin{figure*}
  \centering
  \includegraphics[width=0.48\textwidth]{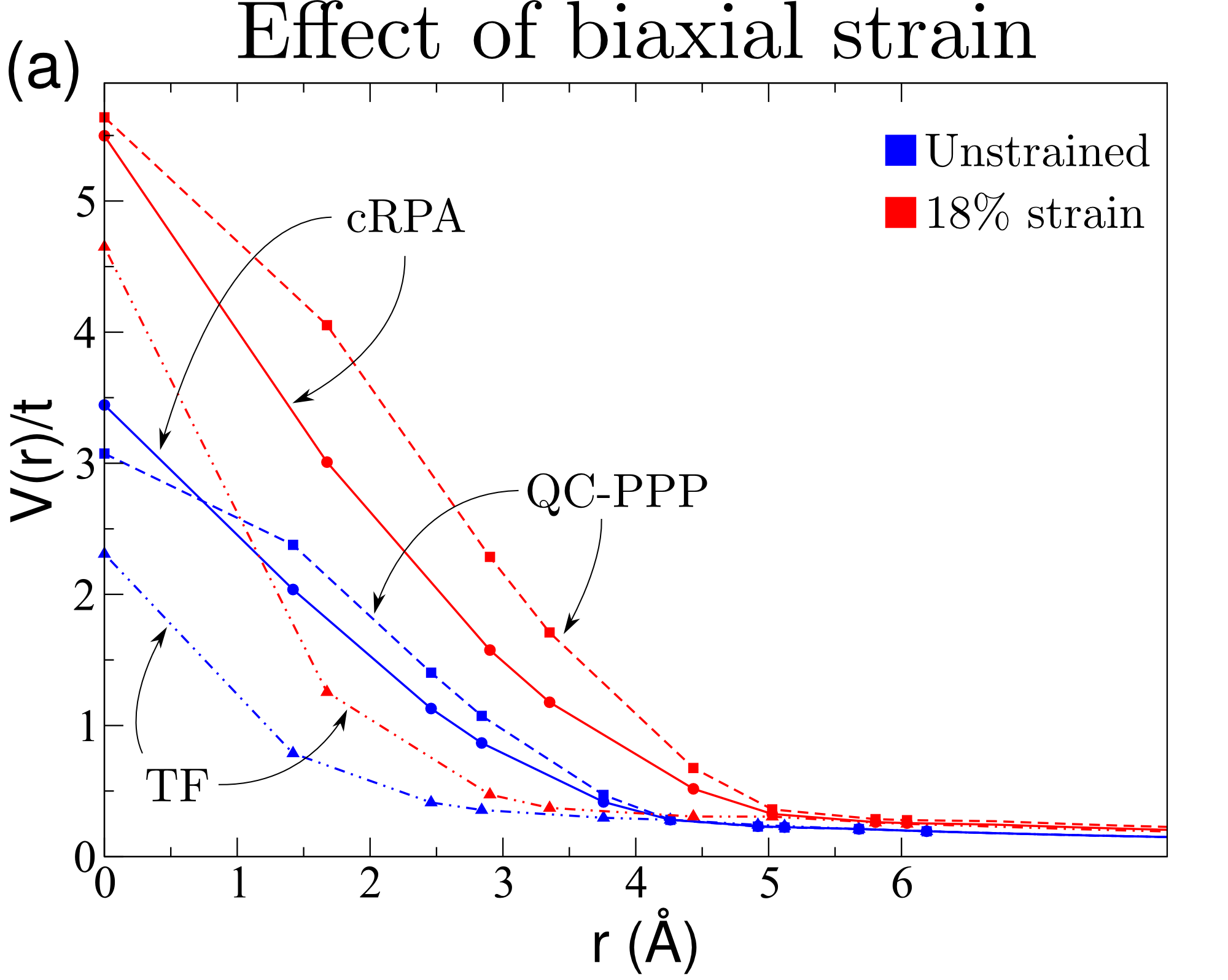}
  \includegraphics[width=0.48\textwidth]{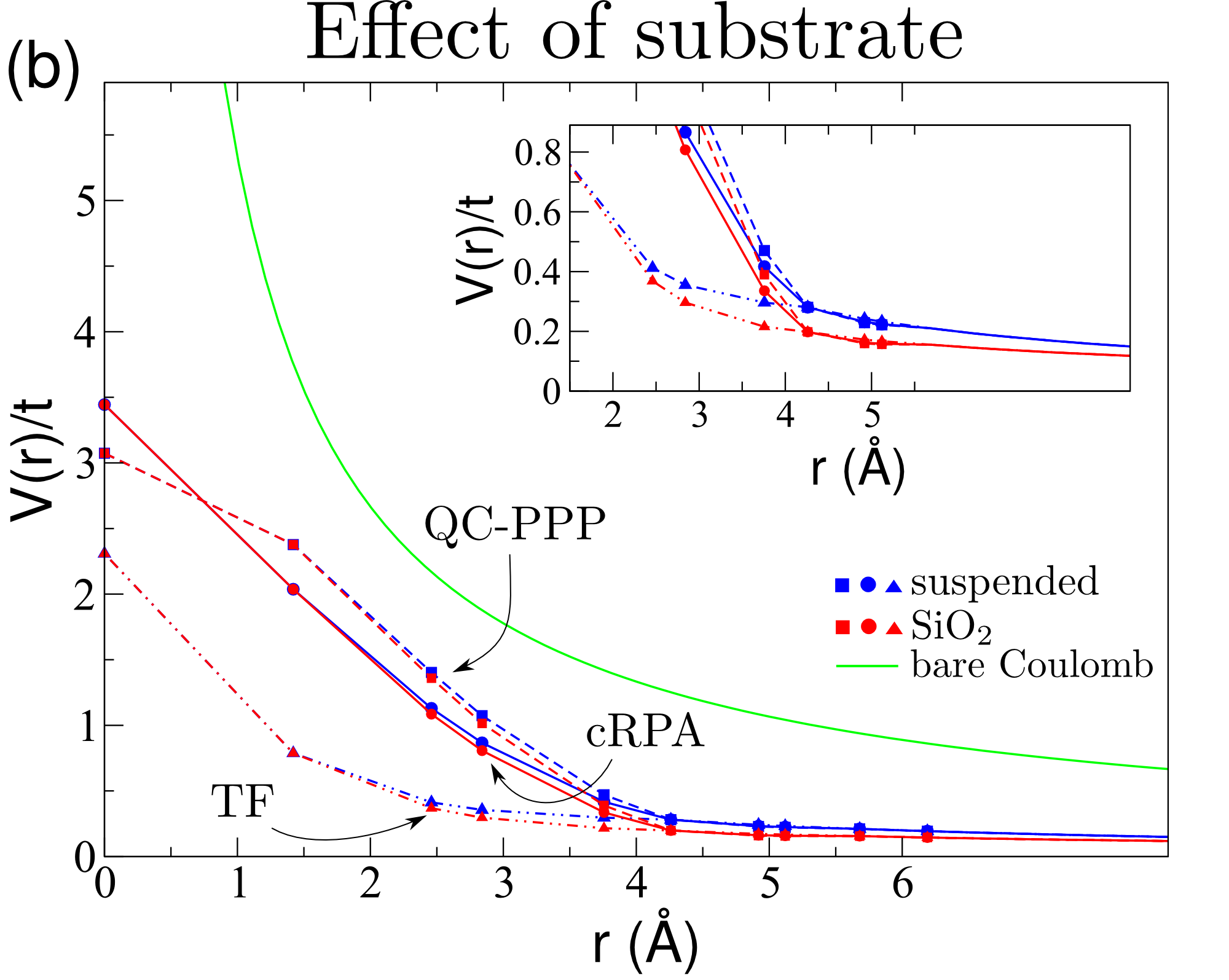}
  \caption{Effect of biaxial strain (left panel) and substrate (right
    panel) on the partially screened Coulomb interaction. We use three
    representative models: constrained Random Phase Approximation
    (cRPA), circles/full curves; quantum chemistry --
    Pariser-Parr-Pople (QC-PPP), squares/dashed curves; and
    Thomas-Fermi (TF), triangles/dot-dashed curves. (a) Suspended
    graphene both unstrained and subject to $18\%$ biaxial strain. (b)
    Unstrained graphene both suspended and deposited on SiO$_{2}$
    compared to the bare Coulomb potential.}
  \label{fig:E-e_IntProfiles}
\end{figure*}

Here we study the possibility to drive graphene across the SM-AFM phase transition by substrate manipulation
or application of biaxial (i.\@ e.\@ uniform and isotropic) strain -- see Fig. \ref{fig:Scheme-Strain}). First we must 
fix the long range Hubbard model's parameters $\mathcal{V}_{i j}$. These are the crucial ingredients determining the 
ground state properties of the system, yet their real values are unknown. We use three representative methods to capture 
the full spatial profile of the partially screened Coulomb interaction for p$_{\textrm{z}}$ electrons in realistic graphene,
and choose $\mathcal{V}_{i j}$ accordingly. These methods will be discussed in detail below, but now we just introduce 
their names: Thomas-Fermi (TF), constrained random phase approximation (cRPA) and the quantum chemistry -- Pariser-Parr-Pople 
(QC-PPP) method. We then investigate the effect of biaxial strain and substrate manipulation on the partially screened Coulomb 
potential $\mathcal{V}(r)$. We find (see Fig. \ref{fig:E-e_IntProfiles}) that biaxial strain strongly modifies the 
$\mathcal{V}(r)$ close to $r=0$ (not affecting the long-range interactions), while changing the substrate (which changes both 
the dielectric screening \cite{Jang-Adam-Fuhrer_PRL:2008} and the amount of disorder \cite{DasSarma_RMP:2011}) only weakly modifies the 
long-range tail of $\mathcal{V}(r)$. Finally, using quantum Monte Carlo techniques (finite temperature 
determinant quantum Monte-Carlo and zero-temperature projective quantum Monte-Carlo), we simulate the ground state of the 
long-range half-filled Hubbard model (in the honeycomb lattice) with the $\mathcal{V}_{i j}$ obtained from $\mathcal{V}(r)$, 
and argue that at least within the Thomas-Fermi approximation, an experimentally feasible \cite{Kim_Nature:2009,Pereira_PRB:2009} 
amount of strain would drive graphene into an interaction driven insulating phase, which could be then measured in 
compressibility, transport or scanning probe experiments.


We now detail the three methods that we use to estimate the partially screened Coulomb interactions that p$_{\textrm{z}}$ 
electrons feel.  (These were chosen since they are very representative of the different approaches that have so 
far been used in the literature). 
The cRPA method (see e.g. Ref.~\cite{Aryasetiawan_PRB:2006}) was adapted 
to graphene by Wehling et al. \cite{Wehling_PRL:2011}. In a systematic way, this method makes use of the electronic
structure of graphene to compute, within the Random Phase 
Approximation, the polarization function $P_{r}$ associated with all the interaction events other than those involving 
two electrons from the $\pi$-bands. Then, the effective (partially) screened Coulomb interaction felt by the p$_{\textrm{z}}$ 
electrons is given by $\mathcal{V}(r) = [1 - V_{bare}(r) P_{r}]^{-1} V_{bare}(r)$, where $V_{bare}(r)$ stands for 
the bare Coulomb potential. The accuracy of this method has long been debated in the literature (see e.g. Ref.~
\cite{Casula_PRL:2012}), and its results are often difficult to interpret physically. For graphene, the fact that 
the Dirac band spans a broad energy window further complicates the application of the cRPA formalism. Notwithstanding 
these difficulties, the cRPA remains the best numerical technique at our disposal to determine the $\mathcal{V}_{i j}$ 
for graphene. In this manuscript we use the cRPA results previously obtained in Ref.~\cite{Wehling_PRL:2011}, 
which compute $U$, $V$ and $t$ for biaxial strains up to $12\%$. In this range of strains all these parameters show a linear 
behavior. In order to obtain the cRPA values of $U$, $V$ and $t$ for up to $18\%$ strain (see Fig. \ref{fig:E-e_IntProfiles}a) 
we have assumed that this linear behavior remains unchanged, extracting $U$, $V$ and $t$ from a linear fit 
to Ref.~\cite{Wehling_PRL:2011}'s numerical results.

The QC-PPP method (see e.g. V\'erges et al. \cite{Verges_PRB:2010}) works by using {\it ab initio} Hartree-Fock and 
post-Hartree-Fock techniques to solve for the ground state energy of molecules comprising a small 
number of benzene rings. These energies are then compared to an exact diagonalization of the long range Hubbard model 
where the Ohno interpolation formula, 
$\mathcal{V}(r) = U / \sqrt{1 + (U r / e^{2})^{2}}$, is assumed 
for the Coulomb interaction. The $\mathcal{V}(0) = U$ is a free parameter that is fixed by requiring the minimization 
of the root-mean square of the ground state energy of the {\it ab initio} calculations and that of the long range 
Hubbard model. The QC-PPP values of $U$ and $V$ used in this manuscript were extracted from Ref.~\cite{Verges_PRB:2010}, which
calculates $\mathcal{V}(r)$ for the phenalenyl (3H-C$_{13}$H$_{9}$) molecule. This method gives an
 an upper bound for the Hubbard $U$ in graphene since larger molecules would have more screening and reduced 
$\mathcal{V}(r)$. Both the validity of the Ohno interpolation and the extrapolation to larger system 
sizes give some reasons for caution. It has nonetheless proven extremely useful for small $\pi$-conjugated planar polycyclic 
aromatic hydrocarbons comprising tens of atoms such as anthracene and polyacenes~\cite{Verges_PRB:2010,footnote1}. 

Finally, inspired by the work of Jung and MacDonald \cite{Jung_PRB:2011} we have constructed a Thomas-Fermi model to account 
for the screening of higher energy bands in graphene. Within the Thomas-Fermi screening approximation the on-site interaction
$U$ is given by
\begin{eqnarray}
  U &=& \frac{e^{2}}{4 \pi \epsilon} \int \textrm{d}^3\mathbf{r}_{1} \textrm{d}^3\mathbf{r}_{2} \, \vert \phi(\mathbf{r}_{1}) 
  \vert^2 \, \frac{e^{-k_{0} \vert \mathbf{r}_{1} - \mathbf{r}_{2} \vert}}{\vert \mathbf{r}_{1}-\mathbf{r}_{2} \vert} \, \vert 
  \phi(\mathbf{r}_{2}) \vert^2 \,, \label{eq:TF-U}
\end{eqnarray}
while the Coulomb interaction between two $\pi$-bands' electrons positioned at neighboring atoms (distance $\delta$) 
$V$ is given by
\begin{eqnarray}
  V &=& \frac{e^{2}}{4 \pi \epsilon} \int \textrm{d}^3\mathbf{r}_{1} \textrm{d}^3\mathbf{r}_{2} \, \vert \phi(\mathbf{r}_{1} 
  + \boldsymbol{\delta}) \vert^2 \, \frac{e^{-k_{0} \vert \mathbf{r}_{1} - \mathbf{r}_{2} \vert}}{\vert \mathbf{r}_{1}
  -\mathbf{r}_{2} \vert} \, \vert \phi(\mathbf{r}_{2}) \vert^2 \,. \nonumber \\ \label{eq:TF-V}
\end{eqnarray}
Here, $\phi(r)$ stands for the p$_{\textrm{z}}$-orbital's wave-function (which we approximate by that of atomic hydrogen). The free
parameter $k_{0}$ in Eqs. (\ref{eq:TF-U}) and (\ref{eq:TF-V}) is fixed by requiring that the hopping integral
\begin{eqnarray}
  t &=& \int \textrm{d}^3\mathbf{r} \, \phi^{*}(\mathbf{r}+\boldsymbol{\delta}) \, \bigg[ - \frac{\hbar^{2} \nabla^{2}}{2 m} 
    + \frac{e^{2}}{4 \pi \epsilon} \sum_{i} \frac{e^{-k_{0} \vert \mathbf{r} - \mathbf{R}_{i} \vert}}{\vert \mathbf{r}-\mathbf{R}_{i} \vert} 
    \bigg] \, \phi(\mathbf{r}) \,, \nonumber \\
\end{eqnarray}
is equal to the literature accepted value of $t_{0} = 2.7$ eV \cite{Reich_PRB:2002}. 
In parallel with what we do for the other two methods, we then interpolate between 
$\mathcal{V}_{i j}$'s short-range values $U$ and $V$ and the long-range tail of $\mathcal{V}_{i j}$ (see below). The procedure
used to compute $\mathcal{V}_{i j}$ of biaxially strained graphene is similar to that discussed earlier~\cite{footnote1}.  

The computationally demanding method employed prevents us from simulating large size systems. In particular, 
one must include the effect of the surrounding electrons since their inter-band polarizability contributes 
at all length scales \cite{DasSarma_RMP:2011} thus modifying the effective dielectric constant from $1/r$ 
to $1/\big[ r (1 + \pi r_{s} /2) \big]$, 
where $r_s = 2 e^2 / \big[ (\kappa_a + \kappa_b) \hbar v_{F} \big]$ is the effective fine structure constant (where $\kappa_a$ 
and $\kappa_b$ are the dielectric constants above and below the graphene flake). The presence of disorder in the 
substrate can also be accounted for by introducing a modified screening function (see e.g. Ref.~\cite{Adam_PRB:2011}).
The full profile of the partially screened Coulomb interaction is obtained by interpolating between the short-range results
, at first neighbor distance, and the long-range tail (assumed to start at the fourth neighbor). 

As we can see in Fig. \ref{fig:E-e_IntProfiles}, the short-range part of the partially screened Coulomb interactions 
$\mathcal{V}(r)$ is strongly affected by biaxial strain (left panel), while its long-range tails are nearly insensitive to 
strain. Manipulating the substrate has a very weak effect on the long-range tails of the partially screened Coulomb 
interactions (right panel).


With the electron-electron interaction profiles of Fig. \ref{fig:E-e_IntProfiles} we have fixed the long range Hubbard 
model's parameters $\mathcal{V}_{i j}$ and explored its ground state using auxiliary field quantum Monte Carlo simulations 
(made possible by recent works \cite{Hohenadler_PRB:2014,Brower_PoSLattice:2011,Ulybyshev_PRL:2014}) -- a numerically exact method 
for investigating strongly correlated systems.  In this manuscript, we use different implementations of the auxiliary 
field quantum Monte Carlo technique: finite temperature determinant quantum Monte-Carlo (DQMC), whose correlation 
functions are given by
\begin{eqnarray}
 \label{Eq:DQMC}
  \langle \hat{O} \rangle &=& \frac{1}{Z} Tr \big[ \hat{O} e^{\beta H} \big] = \frac{1}{Z} \int \mathcal{D}[\phi_{i,\tau}]  
  e^{-S[\phi_{i,\tau}]} \hat{O}[\phi_{i,\tau}] \,,
\end{eqnarray}
(we refer the reader to Ref.~\cite{Santos_BJP:2003} for details);  and zero-temperature projective quantum 
Monte-Carlo (PQMC) (for details see e.g. Ref.~\cite{Assaad_LectureNotes:2002}), where the correlation functions 
are given by 
\begin{eqnarray}
  \label{eq:PQMC}
  \langle \hat{O} \rangle &=& \frac{\langle \Phi_0 \vert \hat{O} \vert \Phi_0 \rangle}{\langle \Phi_0 \vert \Phi_0 
  \rangle} = \lim_{\Theta \rightarrow \infty} \frac{\langle \Psi_T  \vert e^{-\Theta \hat{H}/2} \hat{O} e^{-\Theta \hat{H}/2} \vert 
  \Psi_T \rangle}{\langle \Psi_T \vert e^{-\Theta \hat{H}} \vert \Psi_T \rangle} \,.
\end{eqnarray}
In both cases we use a Hubbard-Stratonovich transformation to convert the interaction 
term into a non-interacting term coupled to an auxiliary field. This transformation enables us to treat 
Hubbard models with non-local electron interactions, provided that the long-range interaction gives rise to a transformation 
matrix that is positive definite (a non-positive definite transformation matrix corresponds to a diverging auxiliary field). 

In particular, we find that the transformation matrix for the case where the $\mathcal{V}_{i j}$ is obtained from 
the QC-PPP method is not positive definite. This is a direct consequence of the interpolation scheme mentioned above, 
which renders the off-diagonal matrix elements associated with the non-local interaction comparable with the diagonal 
elements associated with the local interactions.  As a result, we could not use quantum Monte Carlo to simulate the QC-PPP model.

In the DQMC, we used inverse temperature $\beta=\frac{1}{T}$ in Eq.~\ref{Eq:DQMC} between $24$ and $36$, which is sufficient 
to probe the low-energy behavior of the system. In the PQMC, we chose the Hartree-Fock state as our trial wave-function, $\vert \Psi_T \rangle$, using 
$\Theta t=40$ (see Eq.~\ref{eq:PQMC}) to project the wave-function onto the ground state. We made use of the scaling behavior 
of the antiferromagnetic structure factor $(S_{AFM})$ to estimate the magnetic state of the system. 
\begin{eqnarray}
S_{AFM}=\frac{1}{L^{2}} \sum_{i,j} \Big[\langle m_{i A} m_{j A} \rangle + \langle m_{i B} m_{j B} \rangle \Big] \,,
\end{eqnarray}
where $m_{i C}$ stands for the magnetization of the site located in the atom of sub-lattice $C=A,B$ of the unit cell 
$\mathbf{r}_{i}$, while $L^{2}$ is the number of unit cells (i.e.~$N = 2 L^{2}$ sites). The system's AFM order parameter 
is given by $m_{AFM} = \sqrt{S_{AFM} / (2 L^{2})}$. We have simulated lattice sizes between $L=6$ and $L=15$. In order to take 
finite size effects into account, we utilize $\tilde{m}_{AFM} = m_{AFM} L^{\frac{\beta}{\nu}}$ where we use the critical exponents 
$\beta/\nu\approx 0.9$ (obtained from the best data collapse in Ref.~\cite{Hohenadler_PRB:2014}), compatible with the 
Gross-Neveu universality class \cite{Herbut_PRL:2006,Parisen_PRB:2015}.
\begin{figure}
  \centering
  \includegraphics[width=0.98\columnwidth]{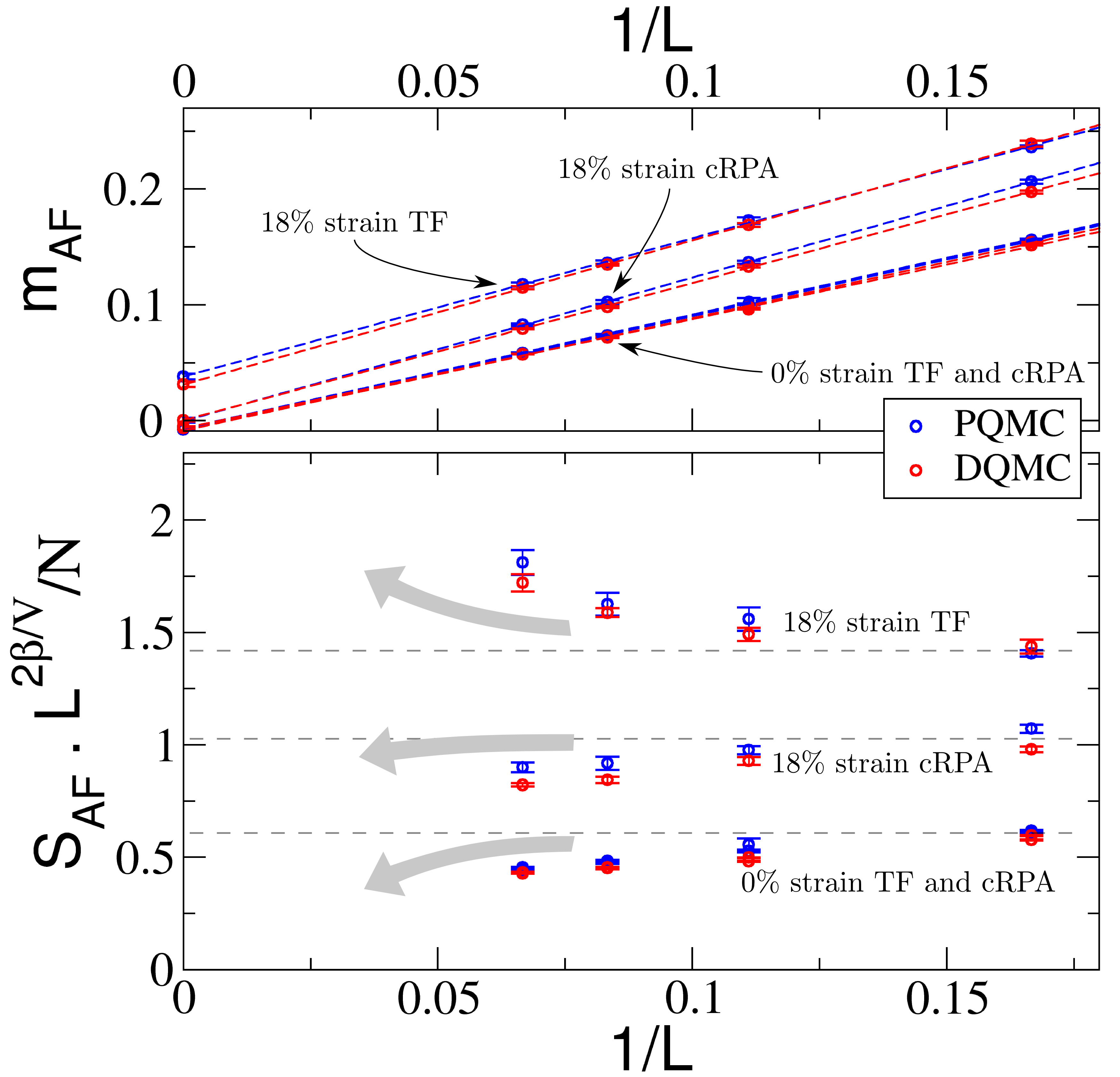}
  \caption{Antiferromagnetic (AFM) order parameter $m_{AFM} =
    \sqrt{S_{AFM} / (2 L^{2})}$ (on top) and scaled antiferromagnetic
    structure factor $S_{AFM} L^{2 \beta/\nu} / N$ (bottom) in
    terms of the inverse system size. We have used both
    projective quantum Monte Carlo (blue) and determinant quantum Monte
    Carlo (red) to study the phase of graphene subject to: $18\%$
    biaxial strain within the Thomas-Fermi model (upper points);
    $18\%$ biaxial strain within the constrained RPA model (middle
    points); $0\%$ strain within both the Thomas-Fermi and the
    constrained RPA models (lower points). We could not simulate the
    quantum chemistry -- Pariser-Parr-Pople model
    with auxiliary field quantum Monte Carlo since its partially
    screened Coulomb potential gives rise to diverging auxiliary
    fields.}
  \label{fig:QMCresults}
\end{figure}

Figure \ref{fig:QMCresults} shows the dependence of $m_{AFM}$ and $\tilde{S}_{AFM}$ with the system size. For unstrained graphene
both the cRPA and TF methods show $\tilde{S}_{AFM}$ decreasing with system size (and $m_{AFM}$ extrapolating to zero in the 
thermodynamic limit $L\to\infty$), indicating that without strain suspended graphene is metallic (in agreement with experimental 
observations). However, most interestingly, with $18\%$ biaxial strain, the TF model shows $\tilde{S}_{AFM}$ increasing with 
the system size (with $m_{AFM}$ extrapolating to a non-zero value when $L\to\infty$), indicating an anti-ferromagnetic
Mott insulator in the thermodynamic limit.  This corresponds to an interaction driven gap of $\Delta = (0.55 \pm 0.05)~{\rm eV}$, 
comparable to estimates in Ref.~\cite{Lee_PRB:2012} obtained by hybrid density functional calculations (Hartree-Fock exchange 
hybridized with generalized gradient approximation for the exchange-correlation) that do not accurately treat 
the effects of electron correlations. Moreover, within the Thomas-Fermi method, our QMC calculations find a critical strain of 
$\eta_{c} \approx 0.15$.  Notice that in this case, $\widetilde{U} = U - V = 3.4 t < U_{c}$ demonstrating that the suggestion 
by Ref.~\cite{Schuler_PRL:2013} for mapping the long range Hubbard model for graphene into an effective onsite Hubbard model 
is quantitatively inaccurate.

Although the TF method has no adjustable parameters, it assumes that the Coulomb interaction between p$_{\textrm{z}}$ 
electrons on the same atom and between neighboring atoms is screened in the same way \cite{Jung_PRB:2011}. This 
assumption slightly overestimates the ratio $U/V$ giving a smaller critical strain $\eta_{c}$ for the SM-AFM transition. On 
the other hand, the canonical cRPA method ignores bandwidth and low-energy spectral weight reduction originated from 
integrating out the high-energy bands \cite{Casula_PRL:2012}. This gives rise to artificially weak partially screened 
Coulomb interactions, resulting in an overestimation of the critical strain $\eta_{c}$. Due to finite sizes, the PPP model 
overestimates the value of $U$ and $V$, and the Ohno interpolation underestimates their difference. However, looking 
at the three models together, we therefore conclude that the profile of the Coulomb potential for realistic graphene 
lies somewhere in between the TF and cRPA estimates. The TF model gives a maximum Mott gap of more than an order of 
magnitude larger than room temperature, and this value sets the upper bound for experiments.


In summary, using the best available models in the literature to estimate the effective Coulomb 
interaction between pz electrons in graphene, we have employed quantum Monte Carlo simulations to 
explore graphene's phase diagram in response to parameters that can be changed experimentally. We 
have found, surprisingly, that manipulating the short-range part of the effective Coulomb potential 
(i.e.~the non-universal and material specific component) is the crucial factor in determining the 
phase of graphene. Most importantly, we show that application of experimentally realistic amounts 
of isotropic strain is a promising route to cross the SM-AFM quantum phase transition and to observe 
a strongly correlated state in this otherwise weakly interacting material.

\begin{acknowledgments}
  HKT thanks Miguel Costa for helpful insights. SA thanks Garnet Chan,
  Jeil Jung and Timo L\"ahde for fruitful discussions. HKT, JNBR and
  SA are supported by the National Research Foundation of Singapore
  under its Fellowship programme (NRF-NRFF2012-01) and by the
  Singapore Ministry of Education and Yale-NUS College through grant
  number R-607-265-01312. PS is supported by the Ministry of Education
  of Singapore through the grant MOE2011-T2-1-108. FFA acknowledges
  the financial support from the DFG grant AS120/9-1. We acknowledge
  the use of the CA2DM and GRC high-performance computing facilities.
\end{acknowledgments}

%

\end{document}